\documentclass[9pt,twocolumn,twoside]{opticajnl}
\journal{opticajournal} 

\setboolean{shortarticle}{false}

\usepackage{siunitx}
\usepackage{setspace}
\usepackage{amsmath}
\usepackage{multirow}
\usepackage{graphicx}

\usepackage{verbatim}
\usepackage{amsfonts}
\usepackage{amssymb}
\usepackage{epstopdf}
\usepackage{dcolumn}
\usepackage{xcolor}
\usepackage{verbatim}
\usepackage{hyperref}
\usepackage{siunitx}
\usepackage[version=4]{mhchem}
\usepackage{textcomp}
\setcitestyle{super} 
\DeclareGraphicsExtensions{.pdf,.eps,.png,.jpg,.mps}

\title{Hertz-Integral-Linewidth Lasers based on Portable Solid-state Microresonators}
\author{Xing Jin$^{1*}$, Xuanyi Zhang$^{1*}$, Fangxing Zhang$^{2\dagger}$, Zhenyu Xie$^{1}$, Shui-Jing Tang$^{1,2,3}$, and Qi-Fan Yang$^{1,2,3\dagger}$\\
$^1$State Key Laboratory for Artificial Microstructure and Mesoscopic Physics and Frontiers Science Center for Nano-optoelectronics, School of Physics, Peking University, Beijing 100871, China\\
$^2$Key Laboratory for Advanced Optoelectronic Integrated Chips of Jiangsu Province, Peking University Yangtze Delta Institute of Optoelectronics, Nantong, Jiangsu 226010, China\\
$^3$Collaborative Innovation Center of Extreme Optics, Shanxi University, Taiyuan 030006, China\\
$^{*}$These authors contributed equally to this work.\\
$^{\dagger}$Corresponding author: leonardoyoung@pku.edu.cn}

\begin{abstract}
Optical reference resonators serve as a cornerstone in various scientific fields. In recent years, there has been an increasing demand for compact ultrastable reference resonators capable of operating in ambient environments, enabling applications beyond the laboratory, such as navigation, portable optical clocks, and remote sensing. Here, we present a compact ultrastable whispering-gallery-mode \ce{MgF2} reference resonator with a high loaded quality factor of $2.24\times 10^9$. The device is packaged in a compact form of 50 × 77 × 90 mm and supports stable optical coupling with polarization-maintaining fiber, which enables robust operation under ambient conditions. Laser stabilization using this resonator yields a phase noise of -105 dBc/Hz at a 10 kHz offset frequency, an integral linewidth of 4 Hz, and a fractional frequency stability of $2.5\times 10^{-14}$ at a 10 ms averaging time. With the high performance and rapid manufacturability, our work offers a promising solution for ultrastable optical frequency references beyond laboratory settings.

\end{abstract}

\setboolean{displaycopyright}{false} 

\begin{document}

\maketitle

\section{Introduction}

Ultrastable lasers are indispensable tools for a wide range of scientific and technological applications, including optical atomic clocks\cite{bloom2014optical}, high-resolution spectroscopy\cite{millot2016frequency}, ultralow-noise photonic microwave synthesis\cite{xie2017photonic,fortier2011generation}, long baseline interferometry\cite{rogers1970very,clivati2020common}, and gravitational wave detection\cite{abbott2016observation}. The generation of ultrastable lasers is typically realized by stabilizing low-noise lasers to ultrastable optical reference resonators, thereby narrowing the laser linewidth and transferring the exceptional stability of the resonators to the laser output\cite{black2001introduction,kondratiev2023recent}. Conventionally, optical reference resonators are based on Fabry–Pérot (F–P) resonators, which offer exceptional noise suppression and frequency stability. State-of-the-art F–P resonators have achieved sub-10-mHz laser linewidths and fractional frequency stabilities below the level of $1 \times 10^{-16}$ \cite{matei20171}. However, these resonators usually require complex fabrication processes that hinder mass production. Additionally, they tend to be bulky and demand stringent operating conditions—such as ultra-high vacuum, precise temperature control, and sophisticated vibration isolation systems—to reach their theoretical stability limits\cite{xie2017photonic,fortier2011generation,he2025highly}. The need for such stringent environmental control not only increases the system’s complexity and cost but also severely restricts its deployment outside a well-controlled laboratory environment. 

In parallel, there is a rapidly growing demand for ultrastable lasers capable of reliable operation in harsh environments, driven by emerging applications such as portable optical atomic clocks\cite{koller2017transportable,roslund2024optical}, geodesy\cite{lisdat2016clock,grotti2018geodesy}, remote sensing\cite{marra2018ultrastable}, navigation\cite{koelemeij2022hybrid}, and miniaturized photonic microwave oscillators\cite{kudelin2024photonic,jin2025microresonator}. These applications often require systems that are not only highly stable but also lightweight, robust, and suitable for deployment on mobile platforms. To meet these requirements, the development of optical reference resonators has increasingly focused on miniaturization, robustness, and portability in recent years \cite{kelleher2023compact}. 

\begin{figure*}[htbp]
	\centering
	\includegraphics[width=\linewidth]{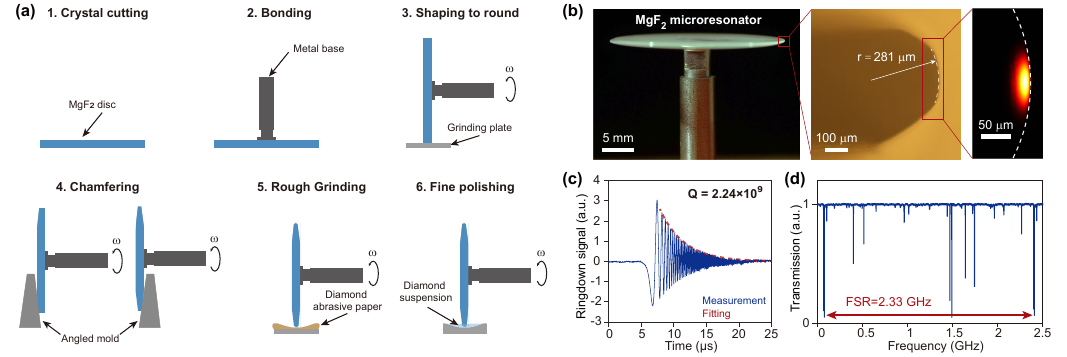}
	\caption{{\bf Fundamental characteristics of the \ce{MgF2} reference resonator.} (a), \ce{MgF2} resonator fabrication process flow. (b), Photograph of the \ce{MgF2} reference resonator (left), microscopic image of its cross section (middle), and the simulated optical mode profile (right). (c), Ringdown measurement of the transmitted optical field when the laser is rapidly swept across the resonance. The exponential decay fitting yields a time constant of 3.9 \textmu s, corresponding to a loaded quality factor of $2.24\times 10^9$. (d) Transmission spectra over a free spectra range (FSR) of the \ce{MgF2} reference resonator.}
	\label{Fig1}
\end{figure*}

For example, miniaturized monolithic F-P resonators enhance mechanical robustness and mitigate environmental sensitivity by integrating the cavity spacer and mirrors into a single structure\cite{zhang2020ultranarrow,zhang2024monolithic}. Laser diodes stabilized to such resonators have demonstrated an integral linewidth of 18 Hz and a fractional frequency stability of $7\times10^{-14}$ at a 10 ms averaging time \cite{zhang2024monolithic}. Whispering-gallery-mode resonators (WGMRs) fabricated from crystalline materials \cite{alnis2011thermal,sprenger2010c,liang2015ultralow} or fused-silica microrods \cite{zhang2019microrod} provide another compact reference platform with high quality (Q) factors. Self-injection-locked laser diodes based on these resonators have reached integral linewidths as low as 30 Hz and fractional frequency stabilities of $3\times10^{-13}$ at a 10 ms averaging time \cite{liang2015ultralow}. Spiral resonators offer a chip-scale optical reference that are compatible with CMOS technology\cite{Lee2014spiral,liu202236}. Such compatibility supports integration with other photonic components and enables scalable mass production. Laser stabilization with these devices via the Pound–Drever–Hall method has yielded an integral linewidth of 36 Hz and a fractional frequency stability of $1.8\times 10^{-13}$ at a 10 ms averaging time \cite{liu202236}. Beyond these monolithic resonators, vacuum-gap F-P resonators employ internally sealed vacuum environments to suppress air-induced refractive index fluctuations, reducing reliance on external ultra-high vacuum systems while maintaining high performance \cite{mclemore2022miniaturizing,liu2024ultrastable,cheng2025harnessing}. These resonators have enabled laser stabilization with fractional frequency stabilities of $6\times10^{-15}$ at a 100 ms averaging time \cite{mclemore2022miniaturizing}.


In this work, we report the development of a compact ultrastable \ce{MgF2} optical reference resonator designed for high-performance laser stabilization. The cross section of the resonator is carefully engineered to suppress excess mode families while preserving a high Q factor. The systematic packaging of the device ensures plug-and-play coupling and enables deployment outside the laboratory within a small volume. Using this resonator under ambient conditions, we demonstrate the stabilization of a commercial fiber laser to achieve outstanding performance. The stabilized laser exhibits a phase noise of –105 dBc/Hz at a 10 kHz offset frequency, an integral linewidth of 4 Hz, and a fractional frequency stability of $2.5\times 10^{-14}$ at a 10 ms averaging time, which is competitive among all the reported laser performances using compact reference platforms.

\section{Device fabrication and characterization}
The \ce{MgF2} reference resonator is fabricated through a meticulous mechanical grinding and polishing process to achieve ultra-high surface quality. As shown in Fig. \ref{Fig1}(a), the fabrication process starts from low-defect, high-purity single-crystal \ce{MgF2} plates, chosen for their exceptionally low optical absorption and minimal thermo-refractive coefficient \cite{savchenkov2007whispering}. The plates are first precisely cut into discs of the required diameter and thickness. These discs are then carefully mounted onto a metal base and holder using UV-curable glue, ensuring mechanical stability and precise alignment for all subsequent processing steps. After that, the mounted \ce{MgF2} disc is first rotated with the metal holder and ground against a horizontal grinding base to form an initial circular profile.
Next, the chamfering process is performed by rotating the disc and grinding it against precision-angled molds to define the desired contour. Surface refinement is subsequently achieved through stepwise mechanical grinding with diamond abrasive papers of progressively decreasing grit size, with microscopic inspection at each stage to monitor improvements in geometric accuracy and surface roughness. The final polishing is carried out in multiple stages using diamond suspensions ranging from 500 nm down to 30 nm in combination with multivacancy pads, followed by optical polishing with cerium oxide and silicon dioxide to eliminate residual surface imperfections. The complete fabrication process takes approximately 8 hours.

\begin{figure*}[t]
	\centering
	\includegraphics{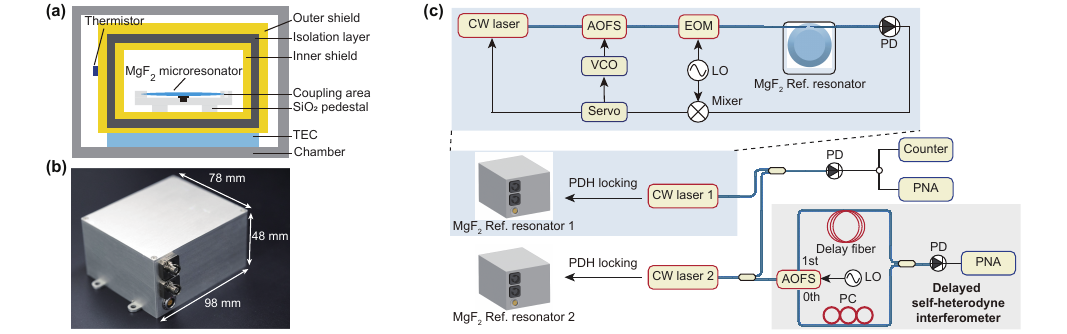}
	\caption{{\bf System packaging of the \ce{MgF2} reference resonator and experimental setup for laser stabilization and performance characterization.} (a), Schematic of the detailed packaging structure of the \ce{MgF2} reference resonator. (b), Photograph of the packaged \ce{MgF2} reference resonator. (c)Experimental setup for laser stabilization and characterization, AOFS: acousto-optic frequency shifter; EOM: electrical optical modulator; PD: photodetector; PC: polarization controller; LO: local oscillator; VCO: voltage-control oscillator; PNA: phase noise analyzer.}
	\label{Fig2}
\end{figure*}

\begin{figure*}[t]
	\centering
	\includegraphics{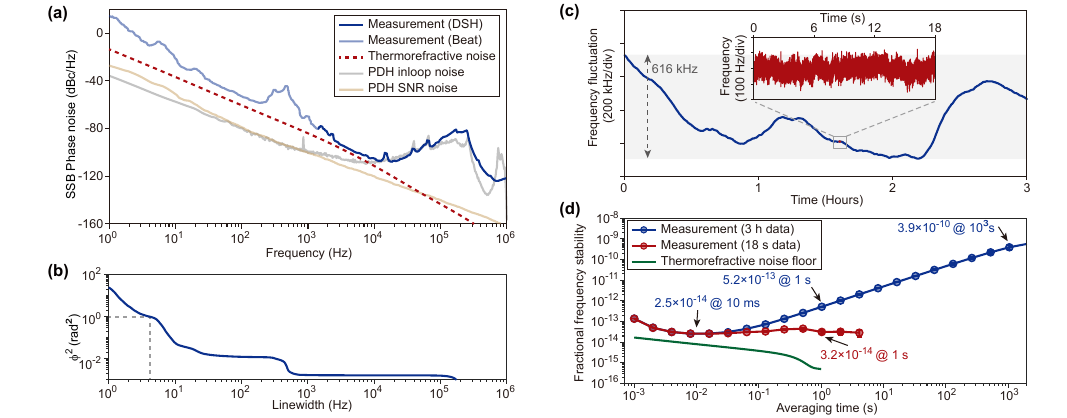}
	\caption{{\bf Phase noise and frequency stability performance of the \ce{MgF2} reference resonator stabilized laser.} (a), Phase noise measurement and simulation. Dark blue: laser phase noise measured with a delayed self-heterodyne (DSH) interferometer; light blue: laser phase noise measured from the beat note between two lasers; red dashed line: simulated thermorefractive noise limit of the resonator; gray: experimentally calibrated in-loop noise; brown: experimentally calibrated signal-to-noise ratio (SNR) noise of PDH locking. (b), Integrated phase noise yields an integral linewidth of 4 Hz, corresponding to an integrated phase variance of 1 rad$^2$. (c), Time trace of the beat note between two lasers independently stabilized to two separate \ce{MgF2} reference resonators. The inset shows the selected 18-second low-drift data of the time trace. (d) Fractional frequency stability calculated from the full 3-hour dataset (blue) and the 18-second segment (red) shown in (c). The green curve represents the calculated thermorefractive noise-limited fractional frequency stability.}
	\label{Fig3}
\end{figure*}
 
As shown in Fig. \ref{Fig1}(b), the fabricated \ce{MgF2} disc resonator features a diameter of approximately 30 mm. The microscopic image of its cross section is displayed in the middle panel of Fig. \ref{Fig1}(b). For further analysis, we approximate the cross section using a circular arc with a curvature radius of 281 \si{\micro\metre}. Finite element simulations based on this geometry model yield an effective mode area of 709 \si{\micro\metre}$^2$. Such a large mode area, combined with the inherently low thermo-refractive coefficient of \ce{MgF2}, contributes to achieving a low thermorefractive noise floor\cite{kondratiev2018thermorefractive}. 

Coupling is achieved using a polarization-maintaining tapered fiber, with the distance and angle precisely controlled by a translation stage and a rotatable fiber clamp to ensure stable excitation of the desired polarization state. For laser stabilization, the system operates in the undercoupled regime to minimize the impact on the loaded Q factor. A continuous-wave fiber laser is coupled into the resonator, and a rapid frequency sweep is performed across one of its resonances. The resulting ringdown trace is shown in Fig. \ref{Fig1}(c). By fitting the decay with an exponential function, we extract a time constant of 3.9 \si{\micro\second}, which corresponds to a high loaded quality factor of $2.24\times10^9$.
Furthermore, we slowly sweep the laser frequency across one free spectral range (FSR) of the resonator and record the transmission spectra. As shown in Fig.\ref{Fig1}(d), only a few optical modes appear within each FSR, which facilitates simultaneous stabilization of two-color lasers to the same mode family of the microresonator. This few-mode characteristic facilitates further noise reduction through the common-mode rejection effect in two-point optical frequency division applications \cite{kudelin2024photonic,jin2025microresonator}.

\section{System package}

To enable user-friendly operation beyond the laboratory, we implement a systematic packaging of the \ce{MgF2} reference resonator. The resonator is securely mounted on a low-thermal-expansion \ce{SiO2} pedestal, providing mechanical stability and minimizing deformation caused by temperature fluctuations. The coupling taper is precisely positioned and fixed using ultraviolet-curable glue to ensure stable and reliable optical coupling.

As depicted in Fig. \ref{Fig2}(a), the entire assembly is enclosed within a multi-layered protective housing. An isolation layer is placed between the inner and outer shields to reduce thermal conduction from the surrounding environment. To further stabilize the thermal environment, a thermistor and thermoelectric cooler (TEC) are integrated into the outer shield, enabling active monitoring and feedback control of the system temperature. This dual approach of passive insulation and active regulation ensures a stable operating point and minimizes resonance frequency drift induced by environmental fluctuations.

As illustrated in Fig. \ref{Fig2}(b), the fully packaged resonator occupies a compact size of only 98 mm$\times$78 mm$\times$48 mm. The module supports direct optical interfacing through an angled physical contact (APC) fiber connector, which provides low-loss coupling and user-friendly plug-and-play operation. Such a robust and portable packaging design enables stable operation of the \ce{MgF2} resonator in ambient conditions, facilitating its practical usage outside conventional laboratory settings.

\section{Laser stabilization and performance characterization}


\begin{table*}
 	\centering
 	\includegraphics[width=\linewidth]{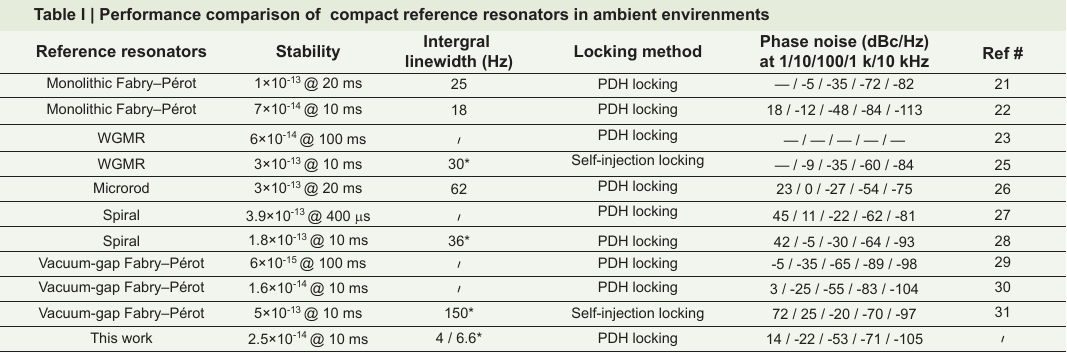}
 	\caption{Linewidth data marked with * correspond to the $1/\pi$ rad$^2$ integral linewidth, all the other values represent the 1 rad$^2$ integral linewidth.}
 	\label{Table}
 \end{table*}

We employed two packaged resonators to stabilize two commertial fiber lasers (NKT 1550 and Precilaser 1550) with closely matched frequencies via the Pound–Drever–Hall (PDH) technique. As shown in the upper panel of Fig. \ref{Fig2}(c), the stabilization scheme adopts a dual-loop feedback architecture. A high-bandwidth feedback loop is implemented through an acousto–optical frequency shifter (AOFS) to increase the locking bandwidth and suppress laser noise over a broad bandwidth. In parallel, a low-bandwidth feedback loop is applied to the piezoelectric transducer of the laser to compensate for slow frequency drifts induced by environmental fluctuations, ensuring long-term stable locking.

One of the stabilized lasers is routed to a delayed self-heterodyne interferometer for phase noise characterization at high offset frequencies (>1.3 kHz) \cite{lao2023quantum}. However, phase noise measurements at low frequencies using this method may be unreliable, as the delayed fiber in the interferometer is susceptible to low-frequency environmental perturbations. To address this, the two stabilized lasers are combined and sent to a photodetector to generate a beat note, which is then analyzed using a phase noise analyzer for low-offset-frequency (<1.3 kHz) characterization. By combining these two methods, we obtain a complete phase noise power spectral density of the stabilized lasers spanning 1 Hz to 1 MHz. Additionally, the beat note is simultaneously monitored with a frequency counter at a 1-ms gate time to evaluate long-term frequency stability.

The phase noise results are shown in Fig. \ref{Fig3}(a), the stabilized laser achieves a phase noise of -105 dBc/Hz at a 10 kHz offset frequency, which is quite near to the simulated thermorefractive noise limit of the reference resonator. The calibrated in-loop noise and the signal-to-noise ratio (SNR) noise \cite{kudelin2024photonic} of the PDH lock are both significantly lower than the measured phase noise at low offset frequencies, owing to the high-Q characteristic of the \ce{MgF2} resonator, which enhances the frequency discriminating sensitivity of the PDH loop. A deviation from the thermorefractive noise limit is observed at offset frequencies below 10 Hz, which is likely attributable to the long-term drift of the resonance frequency. The peak appearing at around 600 Hz might arise from vibrations of the resonator or fluctuations in the coupling taper. By integrating the double-sideband phase-noise power spectral density from 1 MHz to 1 Hz (Fig. \ref{Fig3}(b)), we obtain an integral linewidth of 4 Hz under the 1 rad$^2$ definition and 6.6 Hz under the 1/$\pi$ rad$^2$ definition\cite{hjelme1991semiconductor}.

The long-term stability results are presented in Fig. \ref{Fig3}(c)(d). The beat note of two stabilized lasers exhibits a frequency fluctuation within a 616 kHz range over 3 hours (Fig. \ref{Fig3}(c)). Assuming both lasers possess identical stability, the fractional Allan deviation of each laser derived from the beat note reaches a minimum of $2.5\times 10^{-14}$ at a 10 ms averaging time, nearly approaching the calculated thermorefractive noise floor of the resonator. To further assess the stability limit, we selected an 18-seconds segment of data exhibiting minimal frequency drift (inset of Fig. \ref{Fig3}(c)) for detailed analysis. The calculated stability reaches a minimum of $3.2\times 10^{-14}$ at a 1 s averaging time. The above results demonstrate the exceptional short-term frequency stability achievable with the \ce{MgF2} reference resonators.

\section{Discussion and conclusion}
 
We compare the laser phase noise, integral linewidth, and stability performance with other recently reported results based on compact reference resonators operating in ambient environments, as summarized in Table \ref{Table}. Our results exhibit state-of-the-art phase noise and stability performance among a variety of monolithic reference resonators, including monolithic F-P resonators\cite{zhang2020ultranarrow,zhang2024monolithic}, microrods\cite{zhang2019microrod}, spiral resonators\cite{liu202236,Lee2014spiral}, and WGMRs\cite{alnis2011thermal,liang2015ultralow}. Notably, we demonstrate the first Hertz-level integral linewidth laser based on a compact reference resonator under ambient conditions, a performance not previously reported on any other platforms. Remarkably, the achieved performance is even comparable to that of vacuum-gap F–P resonators\cite{mclemore2022miniaturizing,liu2024ultrastable,cheng2025harnessing}, which typically require more complex fabrication and assembly processes. This combination of high performance, compact form factor, and ease of fabrication highlights the advantages of our approach.

Further improvements in noise performance could be achieved by optimizing the cross section of \ce{MgF2} resonators to enlarge the optical mode volume, thereby achieving a lower thermorefractive noise floor\cite{gorodetsky2004fundamental,matsko2007whispering}. In addition, structural innovations such as sandwich-type composite designs provide an effective approach to compensate for thermal expansion, mitigating long-term resonance frequency drift and enhancing stability over long timescales. These composite structures can be engineered by layering materials with opposite thermal expansion coefficients, allowing the net thermal expansion to be minimized \cite{lim2017chasing}. Beyond resonator design, advances in packaging technology further contribute to performance improvements. For example, incorporating materials with superior thermal insulation or low thermal conductivity into the packaging can enhance thermal shielding, reducing environmental sensitivity. By combining the above strategies, it might be possible to achieve a fractional frequency stability at the $10^{-15}$ level and an integral linewidth approaching the sub-Hertz regime in the future.

In summary, we present a compact reference resonator platform for generating ultralow-noise, highly stable lasers under ambient conditions. Combining Hertz-level integral linewidth, exceptional stability, and rapid fabrication, this platform provides a practical route to portable, field-deployable precision technologies, including portable optical atomic clocks\cite{koller2017transportable,roslund2024optical}, high-accuracy navigation\cite{koelemeij2022hybrid}, geodesy\cite{lisdat2016clock,grotti2018geodesy}, remote sensing\cite{marra2018ultrastable}, and fundamental physics experiments\cite{abbott2016observation}. Our work establishes a versatile foundation for translating laboratory-grade performance into practical, deployable systems, paving the way for future advances across both applied and fundamental science.



\smallskip

\section{Appendix}

{\bf Thermorefractive noise simulation:} 
Thermorefractive noise simulation is based on the fluctuation-dissipation theorem and uses the finite-element approach\cite{kondratiev2018thermorefractive}. In the simulations, the material parameters of \ce{MgF2} are set as follows: refractive index 1.37, material density $3.18\times10^3$ $\SI{}{\kg \ \m^{-3}}$, thermal conductivity $30$ $\SI{}{\W\ \m^{-1}\ \K^{-1}}$, heat capacity $920$ $\SI{}{\J\ \kg^{-1}\ \K^{-1}}$, and thermal-optic coefficient $0.9\times 10^{-6}$ $\SI{}{\K^{-1}}$. The ambient temperature is set to be $300$ $\SI{}{\K}$ in the simulation. 

\noindent {\bf Thermorefractive noise floor calculation:}
The thermorefractive noise floor of the fractional frequency stability is calculated from the simulated thermorefractive phase noise single-sideband (SSB) power spectral density $S_{\varphi}(f)$ using the following method:
\begin{equation}
	\sigma^2(\tau)=4\int_{f_{low}}^{f_{high}}S_{\varphi}(f)f^2\frac{\mathrm{sin}^4(\pi f\tau)}{(\pi f \tau)^2}df
\end{equation}
where $f_{low}=1$ Hz and $f_{high}=10 $ MHz in this calculation. The factor of 4 accounts for the SSB definition of the power spectral density. The corresponding fractional frequency stability is then given by $\sigma (\tau)/\nu_0$ with $\nu_0$ denoting the mode frequency.

\begin{backmatter}
	\bmsection{Funding}
	The project is supported by National Key R\&D Plan of China (Grant No. 2023YFB2806702), Nantong Science and Technology Bureau (MS12022003), and the High-performance Computing Platform of Peking University.
	
	\bmsection{Acknowledgments}
	The authors thank Binbin Nie and Yuanlei Wang for valuable discussions that contributed to the preparation of this manuscript.

	\bmsection{Disclosures}
	The authors declare no conflict of interest.
	
	\bmsection{Data availability} 
	Data underlying the results presented in this article are not publicly available at this time, but may be obtained from the authors upon reasonable request.
\end{backmatter}

\section{References}

\bibliography{ref}

\end{document}